\documentclass[twoside,a4paper,11pt]{sea9}
\usepackage{graphicx}
\usepackage{hyperref}
\usepackage{movie15}
\begin{document}
\pagenumbering{arabic}
\pagestyle{myheadings}
\thispagestyle{empty}
{\flushleft\includegraphics[width=\textwidth,bb=58 650 590 680]{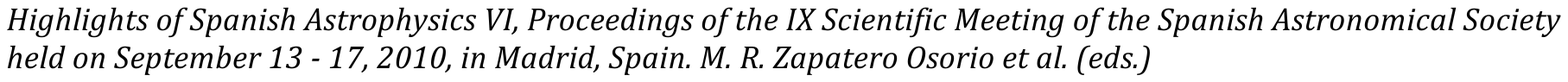}}
\vspace*{0.2cm}
\begin{flushleft}
{\bf {\LARGE
%
%%% TITLE of the paper. 
%%% TITLE of the paper. 
The Calar Alto Observatory: current status and future instrumentation
%
% Do not delete next few lines
}\\
\vspace*{1cm}
%
%%% Include here the LIST OF AUTHORS.
%%% Include here the LIST OF AUTHORS.
%%% Note that the last author has to be preceeded by an AND.
D. Barrado$^{1,2}$,
U. Thiele$^{1}$,
J. Aceituno$^{1}$,
S. Pedraz$^{1}$,
S.F. S\'anchez$^{1}$, 
A. Aguirre$^{1}$,
M. Alises$^{1}$,
G. Bergond$^{1}$,
D. Galad\'\i$^{1}$,
A. Guijarro$^{1}$,
F. Hoyo$^{1}$,
D. Mast$^{1}$,
L. Montoya$^{1}$,
Ch. Sengupta$^{1}$,
E. de Guindos$^{1}$,
and 
E. Solano$^{2}$%
% Do not delete next few lines
}\\
\vspace*{0.5cm}
%
%%% AFFILIATIONS LIST.
%%% and the AFFILIATIONS LIST. Note that one affiliation per line.
%%% Add as many affiliations as necessary. 
$^{1}$
 Centro Astron\'omico Hispano Alem\'an, Calar Alto, (CSIC-MPG),
   C/Jes\'us Durb\'an Rem\'on 2-2, E-04004 Almer\'{\i}a, Spain\\
$^{2}$
Depto. Astrof\'{\i}sica, Centro de Astrobiolog\'{\i}a (INTA-CSIC), ESAC campus, P.O.Box 78, 28691 Villanueva de la Ca\~nada (Madrid), SPAIN \\
%
% Do not delete next few lines
\end{flushleft}
%
% Headings
\markboth{
%%% Type the SHORT version of the paper title.
%%% Type the SHORT version of the paper title.
Calar Alto Observatory: present and future
}{ % Do not delete
%
%%%  First Author \& Second Author   OR   First-author et al. 
%%%  First Author \& Second Author   OR   First-author et al. if the author list 
%%% contains three or more authors.
Barrado et al. 
% 
% Do not delete next few lines
}
\thispagestyle{empty}
\vspace*{0.4cm}
\begin{minipage}[l]{0.09\textwidth}
\ 
\end{minipage}
\begin{minipage}[r]{0.9\textwidth}
\vspace{1cm}
%%%%%%%%%%%%%%%%%%%%%%%%%%%%%%%%%%%%%%%%%%%%%%%%%%%%%%%%%%%%%%%%%
%%%%%%%%%%%%%%%%%%%%%%%%%%%%%%%%%%%%%%%%%%%%%%%%%%%%%%%%%%%%%%%%%
%%%%%%%%%%%%%%%%%%%%%%%%%%%%%%%%%%%%%%%%%%%%%%%%%%%%%%%%%%%%%%%%%
\section*{Abstract}{\small
%
% ABSTRACT ABSTRACT ABSTRACT
% ABSTRACT ABSTRACT ABSTRACT
%%% Type the ABSTRACT of your paper
The Calar Alto Observatory, located at 2168m height above the sea level in continental Europe, holds a significant number of astronomical telescopes and experiments, covering a large range of the electromagnetic domain, from gamma-ray to near-infrared. It is a very well characterized site, with excellent logistics. Its main telescopes includes a large suite of instruments. At the present time, new instruments, namely CAFE, PANIC and Carmenes, are under development. We are also planning a new operational scheme in order to optimize the observatory resources.
%
% Do not delete next few lines
\normalsize}
\end{minipage}
%
%
%%% BODY of the paper
%%% BODY of the paper
%
%%%%%%%%%%%%%%%%%%%%%%%%%%%%%%%%%%%%%%%%%%%%%%%%%%%%%%%%%%%%%%%%%
%%%%%%%%%%%%%%%%%%%%%%%%%%%%%%%%%%%%%%%%%%%%%%%%%%%%%%%%%%%%%%%%%
%%%%%%%%%%%%%%%%%%%%%%%%%%%%%%%%%%%%%%%%%%%%%%%%%%%%%%%%%%%%%%%%%
\section{Introduction \label{section:intro}}

The Calar Alto Observatory is located at 2168m height above the sea level, in the Sierra de los Filabres (Almeria-Spain) at ~45 km from the Mediterranean Sea. It is the second largest European astronomical site in the northern hemisphere just behind the Observatorio del Roque de los Muchachos (located in the island of La Palma), and the most important in the continental Europe (with excellent communications, making logistics easy, unexpensive and reliable). Currently there are six telescopes located in the complex, three of them  directly operated by the Centro Astron\'omico Hispano Alem\'an A.I.E., a partnership between the Spanish National Research Council (CSIC\footnote{www.csic.es}) and the German Max-Plank Society (MPG\footnote{www.mpg.de}). These telescopes include the Zeiss 3.5m, the largest telescope in the continental Western Europe. The observatory is under operations since 1975, when its 1.23 m Zeiss reflector saw first light. The observatory operates a very large array of optical and near-infrared astronomical instrumentation, including imagers and spectrographs with different field-of-view and resolutions.

There has been different attempts to characterize some of the main astronomical properties during its 35 years of operations: (i) Leinert et al. (1995) determined the sky brightness corresponding to the year 1990;  (ii) Hopp \& Fernandez (2002) studied the extinction curve corresponding to the years 1986-2000; (iii) Ziad et al. (2005) estimated the median seeing in the observatory from a single campaign in May 2002; and more recently (iv) S\'anchez et al. (2007) where the optical night sky spectrum was presented, including an analysis of the light pollution, together with a more accurate estimation of the night-sky extinction, the typical seeing, the night-sky brightness and the fraction of useful time; and (v) S\'anchez et al. (2008), where the night sky brightness in the near-infrared and the fraction of useful time was presented. Several of these features are discussed below.

The comprehensive database for the weather is public\footnote{www.caha.es/WDXI/wdxi.php} and it can also be obtained upon request.

%%%%%%%%%%%%%%%%%%%%%%%%%%%%%%%%%%%%%%%%%%%%%%%%%
%%%%%%%%%%%%%%%%%%%%%%%%%%%%%%%%%%%%%%%%%%%%%%%%%  FIGURE 1
%%%%%%%%%%%%%%%%%%%%%%%%%%%%%%%%%%%%%%%%%%%%%%%%%
 \begin{figure}
\center
\includegraphics[width=6.8cm]{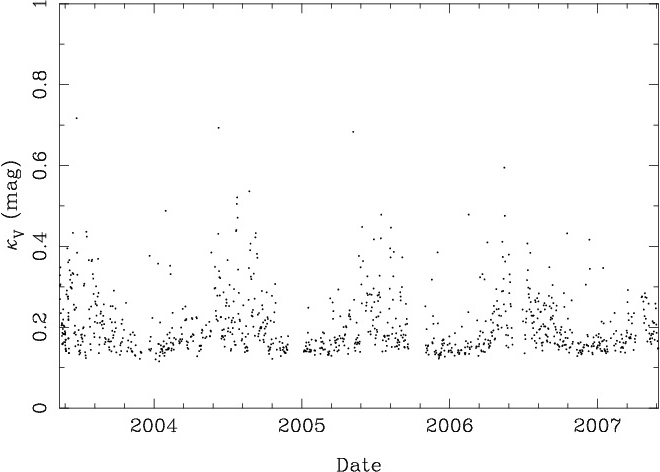} ~
\includegraphics[width=7.5cm]{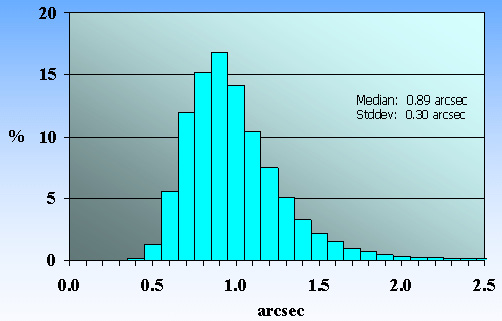} ~
\caption{\label{fig1} 
{\bf Left.-} 
Extinction due to dust at Calar Alto. The values are comparatively smaller than representative values of observatories closer to the Equator, both in the Northern and the Southern Hemispheres. The behaviour is highly seasonal.
{\bf Right.-} 
Seeing distribution (June 2001 -- Sept. 2005) at the Calar Alto Observatory. The median value is about 0.9 arcsec (S\'anchez et al. 2007).
} 
\end{figure}
%%%%%%%%%%%%%%%%%%%%%%%%%%%%%%%%%%%%%%%%%%%%%%%%%
%%%%%%%%%%%%%%%%%%%%%%%%%%%%%%%%%%%%%%%%%%%%%%%%%
%%%%%%%%%%%%%%%%%%%%%%%%%%%%%%%%%%%%%%%%%%%%%%%%%

%%%%%%%%%%%%%%%%%%%%%%%%%%%%%%%%%%%%%%%%%%%%%%%%%%%%%%%%%%%%%%%%%
%%%%%%%%%%%%%%%%%%%%%%%%%%%%%%%%%%%%%%%%%%%%%%%%%%%%%%%%%%%%%%%%%
%%%%%%%%%%%%%%%%%%%%%%%%%%%%%%%%%%%%%%%%%%%%%%%%%%%%%%%%%%%%%%%%%
\section{Site characteristics\label{section:site}}

%%%%%%%%%%%%%%%%%%%%%%%%%%%%%%%%%%%%%%%%%%%%%%%%%%%%%%%%%%%%%%%%%
\subsection{Night-sky spectrum\label{subsection:nightspectrum}}

The optical spectrum at the Calar Alto Observatory shows a strong contamination produced by different  chemicals, in particular from mercury lines, which contribution to the sky-brightness in the different bands is of the order of ~0.09 mag, ~0.16 mag and ~0.10 mag in B, V and R respectively. Regarding the strength of the sodium pollution line in comparison with the air-glow emission, it is very strong, a problem which we expect to address in the near future in collaboration with the regioanl governament and the nearby towns. In any case, the observatory complies with  the IAU recommendations of a dark site. As a matter of fact, CAHA is classified as a Class C site\footnote{\href{http://www.ctio.noao.edu/light_pollution/iau50/prime_dark_sites.htm}{IAU efines as:} ``major observatory sites with operating telescopes having apertures $>$2.5m and  zenith light pollution levels less than the natural variation in night-sky brightness associated with the 11-year solar cycle''}.  A light-pollution regulation was been recently approved by the Andalusian Regional Government, and the observatory is involved in its development, in order to reduce and revert the effect of human light pollution. The effect of such laws have been strong in other astronomical sites, as it was demonstrated for the Kitt-Peak observatory. We expect a similar influence in the next years. Therefore, the darkness of the observatory should improve considerably in the near future, making it even more attractive for new instrumentation.

%%%%%%%%%%%%%%%%%%%%%%%%%%%%%%%%%%%%%%%%%%%%%%%%%
%%%%%%%%%%%%%%%%%%%%%%%%%%%%%%%%%%%%%%%%%%%%%%%%% FIGURE 2
%%%%%%%%%%%%%%%%%%%%%%%%%%%%%%%%%%%%%%%%%%%%%%%%%
 \begin{figure}
\center
\includegraphics[width=7.9cm]{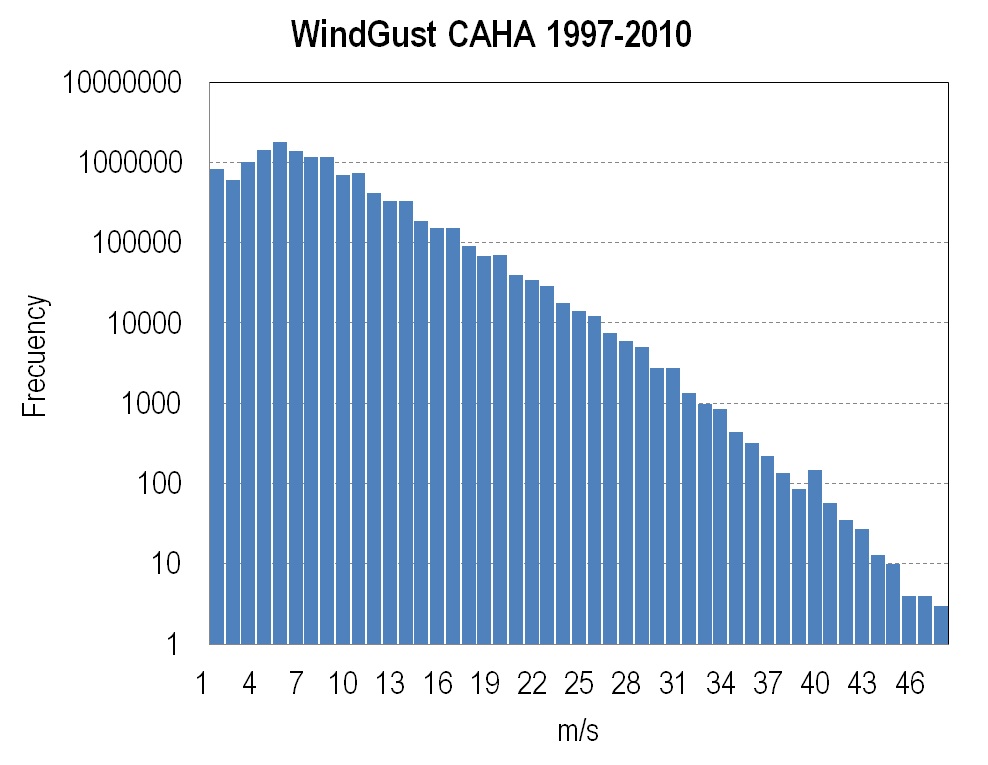} ~
\includegraphics[width=6.8cm]{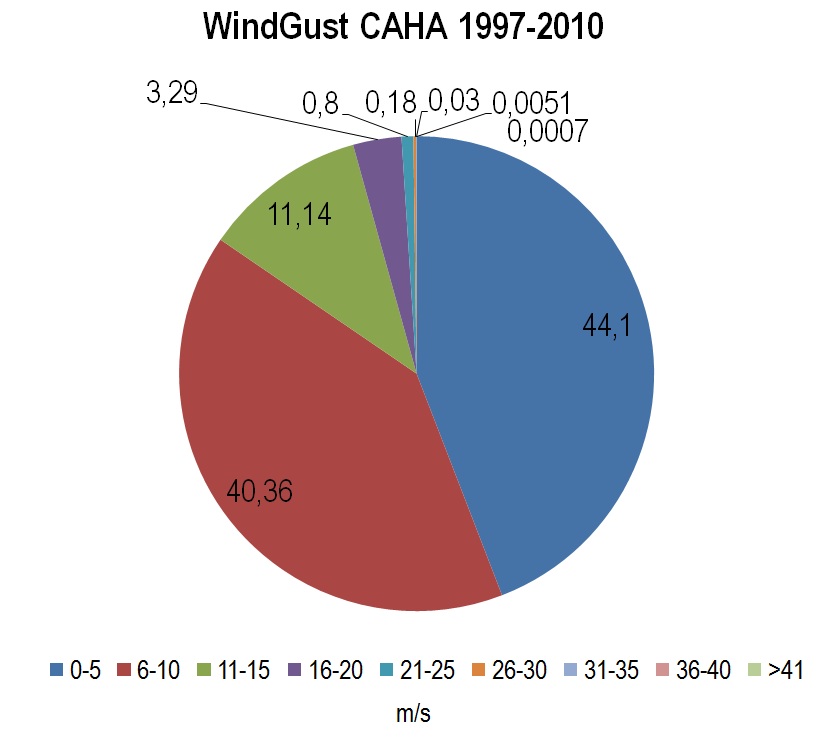} ~
\caption{\label{fig1} 
{\bf Left.-} 
Wind-gust at Calar Alto observatory during the last 13 years. Note the logarithmic scale in the frequency. The absolute velocity maximum is below 47 m/s. 
{\bf Right.-} 
The pie chart  shows the distribution of time spent at different velocities.
} 
\end{figure}
%%%%%%%%%%%%%%%%%%%%%%%%%%%%%%%%%%%%%%%%%%%%%%%%%
%%%%%%%%%%%%%%%%%%%%%%%%%%%%%%%%%%%%%%%%%%%%%%%%%
%%%%%%%%%%%%%%%%%%%%%%%%%%%%%%%%%%%%%%%%%%%%%%%%%

%%%%%%%%%%%%%%%%%%%%%%%%%%%%%%%%%%%%%%%%%%%%%%%%%%%%%%%%%%%%%%%%%
\subsection{Brightness of the night-sky\label{subsection:nightbrightness}}

The zenith-corrected values of the moonless night-sky surface brightness, for a typical dark night, are 22.39, 22.86, 22.01, 21.36, 19.25, 15.95, 13.99, and 12.39 mag arcsec−2 in U, B, V, R, I, J, H and K bands, which indicates that Calar Alto is a particularly dark site for optical and near-IR bands. These values are similar to those at other astronomical sites, including Paranal, La Silla, La Palma and Mauna Kea. Only the last one is clearly darker in the near-IR bands than any of the others.

%%%%%%%%%%%%%%%%%%%%%%%%%%%%%%%%%%%%%%%%%%%%%%%%%%%%%%%%%%%%%%%%%
\subsection{Typical Extinction\label{subsection:extinction}}

The typical extinction in the observatory in the V-band is ~0.15 mag in the winter season, with little dispersion. In summer the extinction has a wider range of values, due to an increase of the aerosol extinction (dust grains), although it does not reach the extreme peaks observed at other sites (Figure 1a). In particular, the extinction in the summer season is much lower than those reported for the observatories in the Canary Islands (Benn \& Ellison 1998), where the influence of the Sahara desert in that archipelago  makes the extinction to rise above 0.25 mag in the V-band for a 20\% of the time in this season. The presence of ozone is negligible at the Calar Alto observatory.

The derived extinction curve shows that the typical extinction in the U-band is ~0.4-0.5 mag, a remarkable good value for observatories at this height.

%%%%%%%%%%%%%%%%%%%%%%%%%%%%%%%%%%%%%%%%%%%%%%%%%%%%%%%%%%%%%%%%%
\subsection{Typical Seeing\label{subsection:seeing}}

The distribution of seeing was derived using the information gathered by the seeing monitor installed in the observatory for more than 4 years (300.000 individual measurements), and the science data from the ALHAMBRA (Moles et al. 2008) survey, from $\sim$5000 individual frames. In both cases it was obtained that the median seeing was about ~0.9", with a sub-arcsecond seeing in  $\sim$70\% of the time (Figure 1b).

%%%%%%%%%%%%%%%%%%%%%%%%%%%%%%%%%%%%%%%%%%%%%%%%%%%%%%%%%%%%%%%%%
\subsection{Wind statistics\label{subsection:wind}}

The historical record in the observatory shows the maximum registered wind speed was 47 m/s in the last 28 years. However, values above 25 m/s are very rare, and these events are smooth and last for short periods of time. Figure 2 displays the frequency of the gust-wind, and the duration for each type of event. As can be seen, there is no sudden variations.  This fact makes the observatory a excellent site for large infrastructures.

%%%%%%%%%%%%%%%%%%%%%%%%%%%%%%%%%%%%%%%%%%%%%%%%%%%%%%%%%%%%%%%%%
\subsection{Useful time for astronomical observations\label{subsection:usefultime}}

The fraction of astronomical useful time at the observatory was derived independently using the information obtained using the time when the extinction monitor is under operation, and the time when the telescopes were open for a long term project (ALHAMBRA, 358 nights distributed along 6 years, Moles et al. 2008). A total of ~70\% of total time was useful to perform astronomical observations under the following criteria: no obvious clouds or cirrus, extinction under 0.2 mag in the V-band, relative humidity under 95\%. The fraction of complete clear nights (more than six continuous hours of useful time), drops to a ~50\%. Finally, a ~30\% of the nights were photometric, defined as nights for which the extinction was stable within the a 20\% of the average value (ie., it was possible to perform direct photometric calibration with an accuracy of 0.2 mags).

%%%%%%%%%%%%%%%%%%%%%%%%%%%%%%%%%%%%%%%%%%%%%%%%%
%%%%%%%%%%%%%%%%%%%%%%%%%%%%%%%%%%%%%%%%%%%%%%%%% FIGURE 3
%%%%%%%%%%%%%%%%%%%%%%%%%%%%%%%%%%%%%%%%%%%%%%%%%
 \begin{figure}
\center
\includegraphics[width=7.9cm]{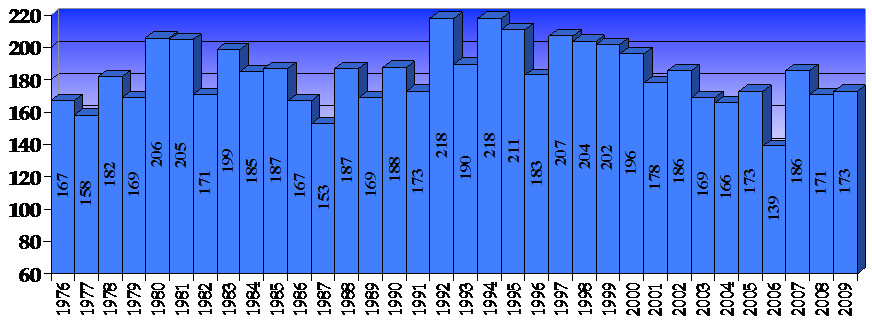} ~
\includegraphics[width=7.1cm]{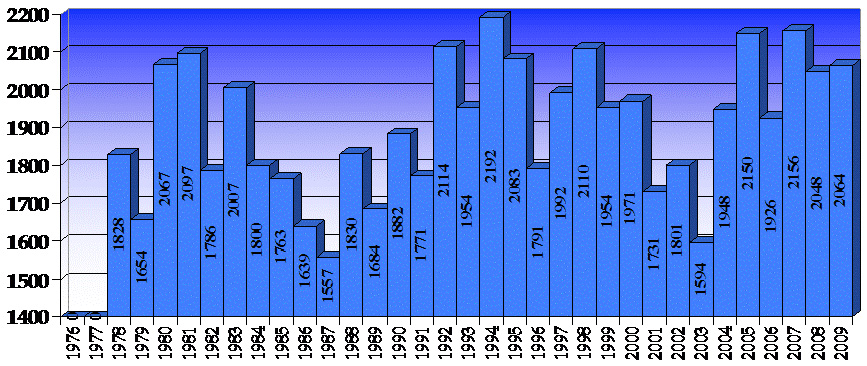} ~
\caption{\label{fig1} 
{\bf Left.-}
Clear nights at Calar Alto Observatory during the period 1976-2009. Clear nights are defined as those having at least 6 observing hours of clear or mostly clear sky (clouds cover not more than 25\%). 
{\bf Right.-}
Total number of hours per year when astronomical observations were carried out, for the period 1976-2009.
} 
\end{figure}
%%%%%%%%%%%%%%%%%%%%%%%%%%%%%%%%%%%%%%%%%%%%%%%%%
%%%%%%%%%%%%%%%%%%%%%%%%%%%%%%%%%%%%%%%%%%%%%%%%%
%%%%%%%%%%%%%%%%%%%%%%%%%%%%%%%%%%%%%%%%%%%%%%%%%

Figure 3ab contains two histograms with the historical record for the  observatory (for the period 1976-2009). The first one represents the number of clear nights (defined as those having at least 6 observing hours of clear or mostly clear sky, with clouds cover below 25\%). The data collected during these  30 years (about three solar activity cycles) clearly indicate that the number of clear nights are around 170 nights per year. See second histogram displays the number of useful hours per year, a more representative quantity since it is closely connected with the observations and the amount of  data which can be produced. The histogram is highly variable, but during the last years an average of 2000 hours has been achieved. This fact might be related with the extensively use of service mode, much more effective than programs executed under a conventional visitor mode. Assuming 9 useful hours per night, 2000 hours are equivalent to 222 full nights per year.

%%%%%%%%%%%%%%%%%%%%%%%%%%%%%%%%%%%%%%%%%%%%%%%%%
%%%%%%%%%%%%%%%%%%%%%%%%%%%%%%%%%%%%%%%%%%%%%%%%% FIGURE 4
%%%%%%%%%%%%%%%%%%%%%%%%%%%%%%%%%%%%%%%%%%%%%%%%%
 \begin{figure}
\center
\includegraphics[width=6.9cm]{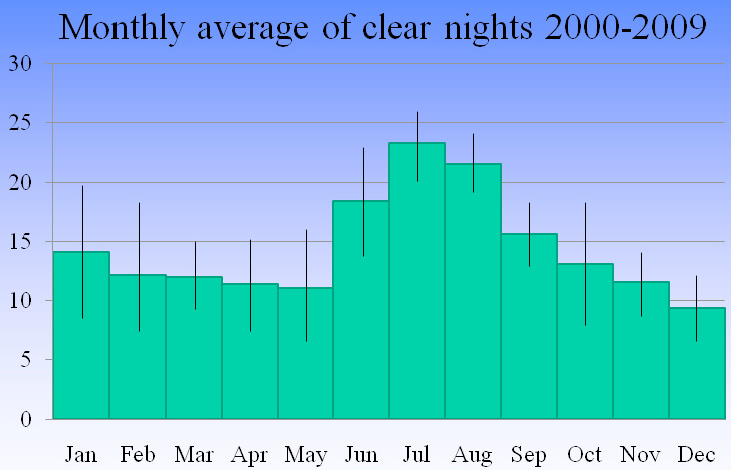} ~
\includegraphics[width=7.0cm]{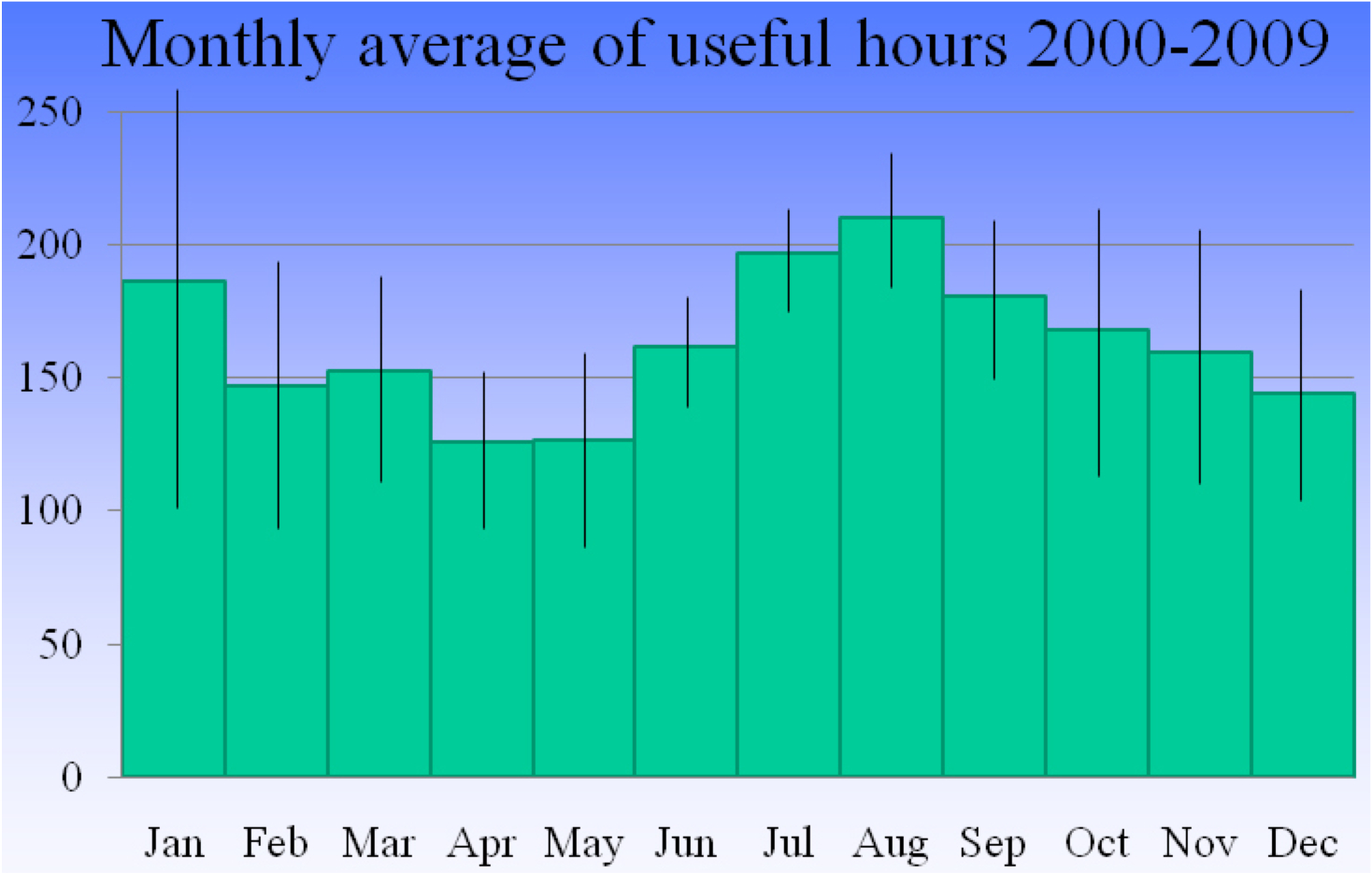} ~
\caption{\label{fig1} 
{\bf Left.-} 
Average for the number of clear nights per month during the period 2000-2009. The bars provide an  idea of the range of variability for each month.
{\bf Right.-}
 Monthly average for the number of useful hours for the period 2000-2009.} 
\end{figure}
%%%%%%%%%%%%%%%%%%%%%%%%%%%%%%%%%%%%%%%%%%%%%%%%%
%%%%%%%%%%%%%%%%%%%%%%%%%%%%%%%%%%%%%%%%%%%%%%%%%
%%%%%%%%%%%%%%%%%%%%%%%%%%%%%%%%%%%%%%%%%%%%%%%%%

As in the case of the extinction (subsect.~\ref{subsection:extinction}), the effectiveness of the observatory, in terms of nights or hours per month, is highly seasonal. Figure 4 illustrates this fact. In the first panel, we represent the number of clear nights per month. There is a conspicuous peak during the summer and a minimum in December, but January shows a  very large dispersion, with years containing 20 clear nights and other with less than 10 clear nights. When expressed in terms of useful hours (Figure 4b) the situation is more balanced. The summer peak is still present, but modulated by the night length. Therefore, the minimum is located during early Spring. Surprisingly, January can be either very poor in useful hours or outstanding.

%%%%%%%%%%%%%%%%%%%%%%%%%%%%%%%%%%%%%%%%%%%%%%%%%
%%%%%%%%%%%%%%%%%%%%%%%%%%%%%%%%%%%%%%%%%%%%%%%%% TABLE 1
%%%%%%%%%%%%%%%%%%%%%%%%%%%%%%%%%%%%%%%%%%%%%%%%%
\begin{table}[ht] 
\caption{Current and future instrumentation} 
%\center
%\begin{minipage}{0.5\textwidth}
%\center
\begin{tabular}{c|cc|cc} 
\hline\hline 
Telescope & Optical  & Near-IR  & Optical & Near-IR     \\ 
          &  2010    & 2010     & 2014    & 2014        \\ 
\hline 
 3.5m     & PMAS     & Omega2000& PMAS    & Carmenes$^3$\\ 
          & TWIN     & OmegaCass& (TWIN)  & (PANIC)     \\ 
          & LAICA    &          &         &  \\ 
          & MOSCA    &          &         &  \\ 
\hline 
 2.2m     & CAFOS    & MAGIC    & CAFE    & PANIC \\ 
          & BUSCA    & PANIC$^2$& (CAFOS) &  \\ 
          & Astralux &          &         &  \\ 
          & CAFE$^1$ &          &         &  \\ 
\hline 
 1.23m    & CCD camera &        & CCD camera &  \\ 
\hline
\end{tabular} 
\\
$^1$ Commissioning 2011.\\
$^2$ Commissioning 2012.\\
$^3$ Commissioning 2014.\\
Additional information at: \\
www.caha.es/telescopes-overview-and-instruments-manuals.html
%\end{minipage}
\label{tab1} 
\end{table}
%%%%%%%%%%%%%%%%%%%%%%%%%%%%%%%%%%%%%%%%%%%%%%%%%
%%%%%%%%%%%%%%%%%%%%%%%%%%%%%%%%%%%%%%%%%%%%%%%%% TABLE 1
%%%%%%%%%%%%%%%%%%%%%%%%%%%%%%%%%%%%%%%%%%%%%%%%%

%%%%%%%%%%%%%%%%%%%%%%%%%%%%%%%%%%%%%%%%%%%%%%%%%%%%%%%%%%%%%%%%%
\subsection{The protection of the sky quality\label{subsection:skyprotection}}

The regional government of Andalusia (Junta de Andaluc\'{\i}a) approved in 2007 an environmental law that includes a chapter on light pollution. This law becomes fully effective after the publication of its ancillary regulation (August 2010). Calar Alto has been advising Junta de Andaluc\'{\i}a from the beginning of this legal project, with one representative in the advisory committee established to define the general terms of the law and its regulation. The law foresees creating an Office for the Protection of Night Sky against Light Pollution, and Calar Alto has secured its participation in this official institution. A protocol is under preparation and its implementation  will allow CAHA to perform specific and individual negotiations with all the relevant cities and towns in the neighbourhood of the observatory.  Therefore, the quality as an outstanding astronomical site will be secured for many years to come.

%%%%%%%%%%%%%%%%%%%%%%%%%%%%%%%%%%%%%%%%%%%%%%%%%%%%%%%%%%%%%%%%%
%%%%%%%%%%%%%%%%%%%%%%%%%%%%%%%%%%%%%%%%%%%%%%%%%%%%%%%%%%%%%%%%%
%%%%%%%%%%%%%%%%%%%%%%%%%%%%%%%%%%%%%%%%%%%%%%%%%%%%%%%%%%%%%%%%%
\section{Current operations and future instrumentation}
 
The observatory operates three telescopes at the present time, with primary mirrors of 3.5m 2.2 and 1.23 meters. The instrumental suite is very diverse in the first two cases, and include optical and near-infrared imagers and spectrographs (six instruments for the 3.5m and four for the 
2.2m\footnote{See: www.caha.es/telescopes-overview-and-instruments-manuals.html}). In order to optimize the observatory resources (man-power, financial and  observing time), we are   introducing  a number of changes in  our operational mode, including the implementation of a public archive and a simplified suite of instruments, following the recommendations of the European Telescopes Strategic Review Committee (ETSRC), which was  appointed by the ASTRONET Board in coordination with the OPTICON Executive Committee.  The current and possible future suite of instruments are summarized in Table 1.

%%%%%%%%%%%%%%%%%%%%%%%%%%%%%%%%%%%%%%%%%%%%%%%%%
%%%%%%%%%%%%%%%%%%%%%%%%%%%%%%%%%%%%%%%%%%%%%%%%% FIGURE 5
%%%%%%%%%%%%%%%%%%%%%%%%%%%%%%%%%%%%%%%%%%%%%%%%%
 \begin{figure}
\center
\includegraphics[width=15.2cm]{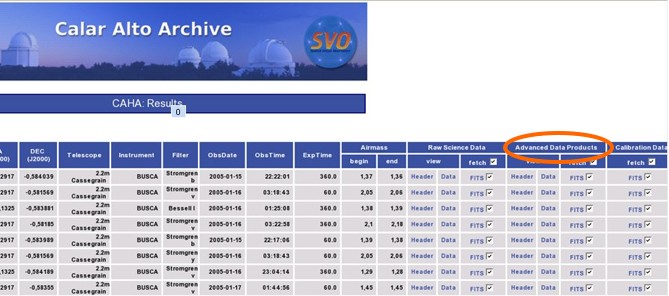} ~
\caption{\label{fig1} 
An example of a query in the CAHA public archive, developed under VO standards.} 
\end{figure}
%%%%%%%%%%%%%%%%%%%%%%%%%%%%%%%%%%%%%%%%%%%%%%%%%
%%%%%%%%%%%%%%%%%%%%%%%%%%%%%%%%%%%%%%%%%%%%%%%%%
%%%%%%%%%%%%%%%%%%%%%%%%%%%%%%%%%%%%%%%%%%%%%%%%%

%%%%%%%%%%%%%%%%%%%%%%%%%%%%%%%%%%%%%%%%%%%%%%%%%%%%%%%%%%%%%%%%%
\subsection{The CAHA public archive}

In collaboration with the Center of Astrobiology (CAB), which hosts the Spanish node of the Virtual Observatory\footnote{Spanish Virtual Observatory: svo.cab.inta-csic.es/} (Solano 2006), we have developed a web-based tool in order to provide  access to the data acquired at Calar Alto Observatory. All data taken after July 1st 2010 will be made public after one year of proprietary rights. For data collected prior that date (starting January 2008), we have requested permit from the principal investigators for each individual program. The final goal is to provide raw and reduced data, as well as meta-data and links to other VO-compliant archives and tools (as an example, see VOSA, Bayo et al. 2009), in order to maximize the scientific exploitation of the data. Figure 5 includes an example of a CAHA Public Archive query. We expect that the utility will be opened in early 2011.

%%%%%%%%%%%%%%%%%%%%%%%%%%%%%%%%%%%%%%%%%%%%%%%%%%%%%%%%%%%%%%%%%
\subsection{Legacy programs}

The observatory has recently approved its first Legacy program, the The CALIFA\footnote{Additional information at www.caha.es/sanchez/legacy/oa/index.php?center=members.php} survey (Calar Alto Legacy Integral Field Area Survey, S\'anchez et al. 2010, 2011; Marino et al. 2010). This program will provide
the largest and most comprehensive wide-field IFU survey of galaxies carried out to date,
addressing several fundamental issues in galactic structure and evolution. We will observe a
statistically well-defined sample of $\sim$600 galaxies in the local universe using 210 observing
nights with the PMAS/PPAK integral field spectrophotometer, mounted on the
Calar Alto 3.5m telescope. The defining science drivers for the project are: a) star formation
and chemical history of galaxies, b) the physical state of the interstellar medium, c)
stellar and gas kinematics in galaxies, and d) the influence of the AGNs on galaxy evolution.
 CALIFA will provide a valuable bridge between large single-aperture
surveys such as SDSS and more detailed studies of individual galaxies with PPAK and other instruments.

Additional Legacy program might be implemented in the future, specially at the 2.2m telescope.

%%%%%%%%%%%%%%%%%%%%%%%%%%%%%%%%%%%%%%%%%%%%%%%%%
%%%%%%%%%%%%%%%%%%%%%%%%%%%%%%%%%%%%%%%%%%%%%%%%% FIGURE 6
%%%%%%%%%%%%%%%%%%%%%%%%%%%%%%%%%%%%%%%%%%%%%%%%%
 \begin{figure}
\center
\includegraphics[width=7.5cm]{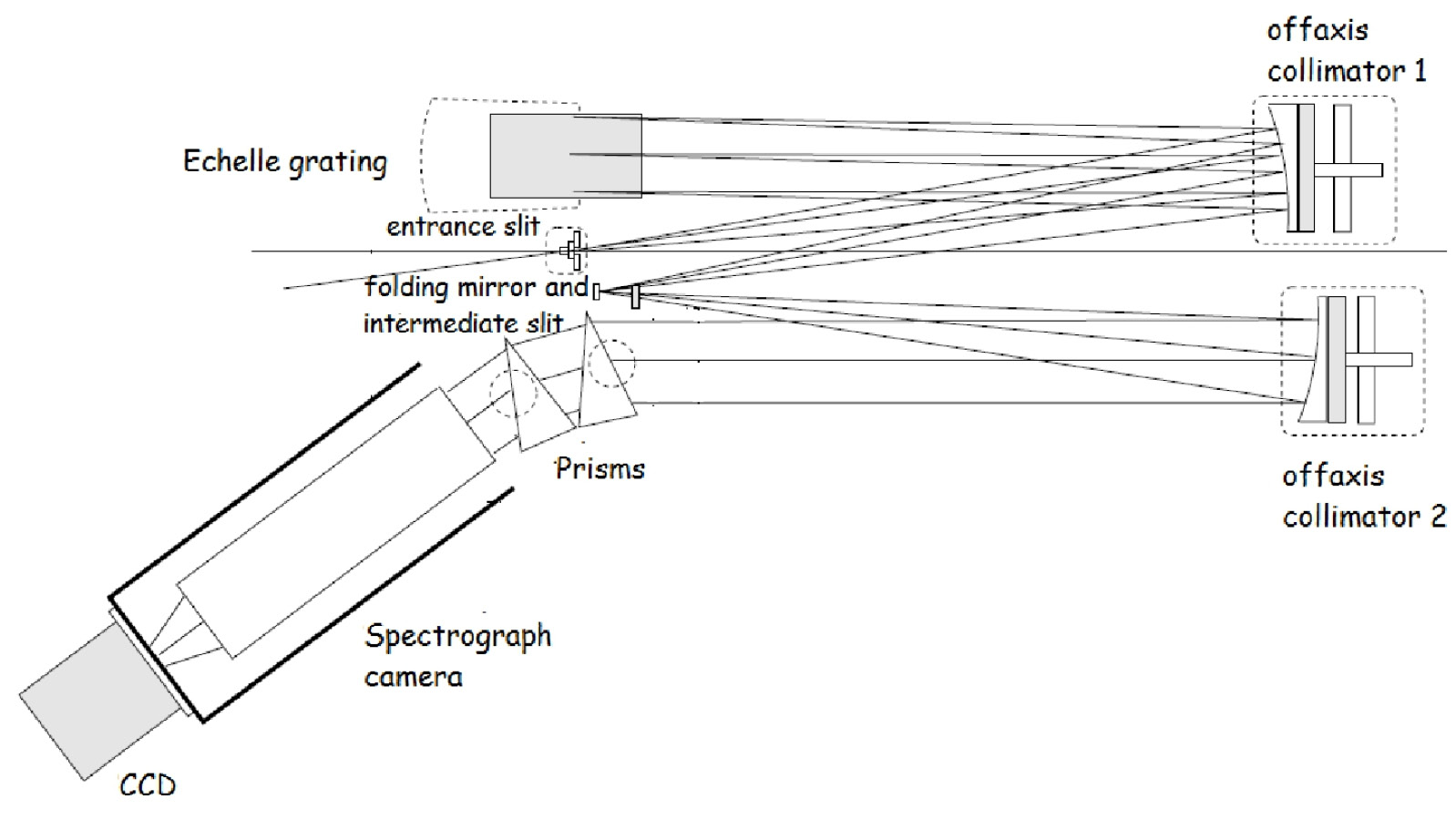} ~
\includegraphics[width=6.3cm]{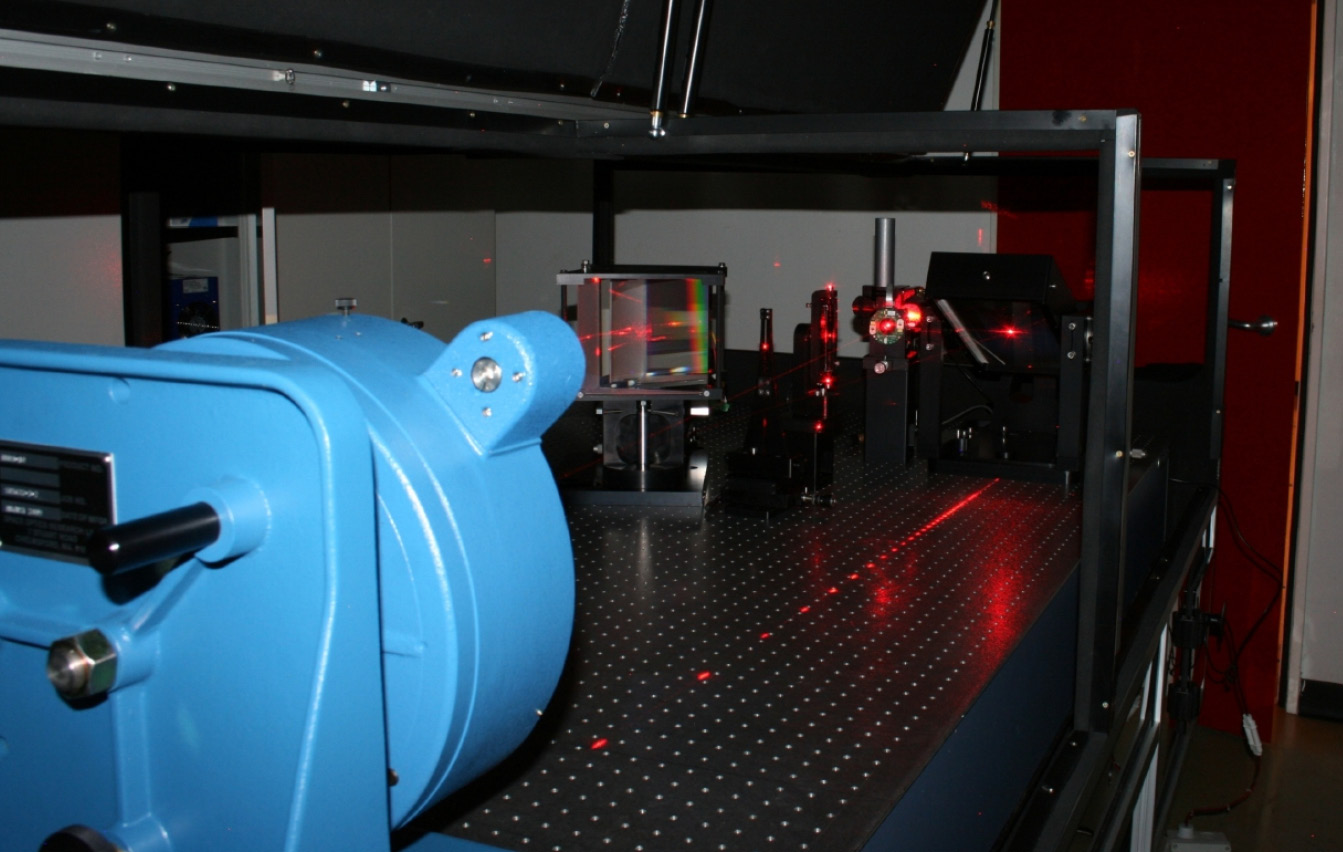} ~
\caption{\label{fig1} 
{\bf Left.-} 
Scheme of the optical design of CAFE, including its main components.
{\bf Right.-}
First assembling of CAFE of its optical bench, in September 2010. The
laser is used to align the different components.
} 
\end{figure}
%%%%%%%%%%%%%%%%%%%%%%%%%%%%%%%%%%%%%%%%%%%%%%%%%
%%%%%%%%%%%%%%%%%%%%%%%%%%%%%%%%%%%%%%%%%%%%%%%%%
%%%%%%%%%%%%%%%%%%%%%%%%%%%%%%%%%%%%%%%%%%%%%%%%%

%%%%%%%%%%%%%%%%%%%%%%%%%%%%%%%%%%%%%%%%%%%%%%%%%%%%%%%%%%%%%%%%%
\subsection{New instrumentation}

At the present time, we are developing three new instruments, namely a high-spectral resolution optical spectrograph for general use at the 2.2m (CAFE), a wide field-of view (30x30 arcmin) near-infrared camera for the 2.2m and the 3.5m (PANIC), and a high-spectral near-infrared spectrograph for exoplanetary searches (Carmenes). More information can be found  in these proceedings or below for CAFE.

The Calar Alto Fiber-fed Echelle spectrograph (CAFE) is an instrument under construction
at CAHA to replace FOCES, the high-resolution echelle spectrograph at the 2.2m Telescope
of the observatory. FOCES is a property of the Observatory of the Munich University, and
it was recalled  from Calar Alto by this institution in 2009. 
The use of this instrument was very extensive, and represented  a substantial fraction
of the 2.2m telescope time during its operational life-time. Due to this fact, the observatory 
decided in 2008 to to build an improved replacement.

CAFE share its basic characteristics with those of FOCES. However, significant improvements
have been introduced in the original design (adding new calibration units), the quality
of the materials, and the overall stability of the system. It is expected that the overall efficiency and the quality of the data will be significantly improved with respect to its predecessor.
In particular, CAFE is design and built to achieve resolutions of R$\sim$70000, which will be
kept in the final acquired data, allowing it to compete with current operational extrasolar
planets hunters.

%%%%%%%%%%%%%%%%%%%%%%%%%%%%%%%%%%%%%%%%%%%%%%%%%
%%%%%%%%%%%%%%%%%%%%%%%%%%%%%%%%%%%%%%%%%%%%%%%%% TABLE 2
%%%%%%%%%%%%%%%%%%%%%%%%%%%%%%%%%%%%%%%%%%%%%%%%%
\begin{table}[ht] 
\caption{Comparison between FOCES and CAFE} 
\center
\begin{minipage}{0.9\textwidth}
\center
\begin{tabular}{lcc} 
\hline\hline 
Parameter             & FOCES                & CAFE\\
\hline 
Design                & Echelle spectrograph & Echelle spectrograph \\
Telescope             & Calar Alto 2.2m      & Calar Alto 2.2m      \\
Operational live-time & 1997-2009            & Starting in 2011     \\
Resolution            & 46000/64000          & 70000                \\
CCD pixel scale       & 24/15$\mu$m          & 13.5$\mu$m           \\
CCD Cooling system    & Liquid nitrogen      & Peltier              \\
Wavelength range      & 3800-7450$\AA$$^1$   & 3800-7450$\AA$       \\
Stability system      & Un-stabilized        & Vibrations and Temperature\\
Moving parts          & Slit,Grating and Prisms & No moving parts\\
Calibration system    & Th-Ar lamps          & Th-Ar lamps and Iodine-cell\\
Quality of optical components & $\lambda$/10 & $\lambda$/20\\
Expected S/N$^2$      & $\sim$100            & $>$150 \\
\hline
\multicolumn{3}{l}{(1) Variable up to 9400$\AA$.}\\
\multicolumn{3}{l}{(2) For a G-star of V=10 mag, 1h exposure time.}\\
\end{tabular} 
\end{minipage}
\label{tab1} 
\end{table}
%%%%%%%%%%%%%%%%%%%%%%%%%%%%%%%%%%%%%%%%%%%%%%%%%
%%%%%%%%%%%%%%%%%%%%%%%%%%%%%%%%%%%%%%%%%%%%%%%%% TABLE 2
%%%%%%%%%%%%%%%%%%%%%%%%%%%%%%%%%%%%%%%%%%%%%%%%%

It is expected that the final assembling of the instrument take place before the end of 2010.  Telescope commissioning is expected for spring 2011.

We expect that PANIC will be operational in 2012 and Carmenes, which has received green light very recently, in 2014.

%%%%%%%%%%%%%%%%%%%%%%%%%%%%%%%%%%%%%%%%%%%%%%%%%%%%%%%%%%%%%%%%%
%%%%%%%%%%%%%%%%%%%%%%%%%%%%%%%%%%%%%%%%%%%%%%%%%%%%%%%%%%%%%%%%%
%%%%%%%%%%%%%%%%%%%%%%%%%%%%%%%%%%%%%%%%%%%%%%%%%%%%%%%%%%%%%%%%%
\section{Summary}

The Calar Alto observatory has two main properties: a very well-characterized and excellent astronomical site (specially for spectroscopy) and outstanding logistics (excellent communications and a location in continental Europe). The current and future suite of instruments for the optical and near-IR telescopes, and the support of two of the main research institutions in Europe
(the Spanish National Research Council and the German Max-Plank Society, channeled through the Instituto de Astrof\'{\i}sica de Andaluc\'{\i}a and the Max-Plank Institut f\"ur Astronomie), provide the necessary conditions for its continuation as relevant astronomical facility.

%
%
% Do not delete the next line
\small  % Do not delete
%
%%% Comment the following line if you do not have acknowledgments.
\section*{Acknowledgments}   % Do not delete if you declare acknowledgments
%
%%% ACKNOWLEDGMENTS
%%% ACKNOWLEDGMENTS

We thank the {\it ``Viabilidad , Dise\~no , Acceso y Mejora'' } funding programs,
ICTS-2008-24 and ICTS-2009-10, of the Spanish Ministry of Science, and the
{\it Proyecto de Excelencia} funding program {FQM-08-00360} of the {\it Junta
  de Andaluc\'{\i}a}, for the support given to this project. 
The observatory will not be possible without the superb professionalism of its staff.

%
% Do not delete the next few lines

%
\end{document}